\documentclass[showpacs,groupedaddress,secnumarabic,titlepage,nobibnotes,twocolumn,showkeys,aps]{revtex4}
\usepackage{amssymb}
\usepackage{graphicx}
\usepackage{dcolumn}
\usepackage{bm}
\usepackage{rotating}

\begin{document}

\title{Electron Mobility and Magneto Transport Study of Ultra-Thin Channel
Double-Gate Si MOSFETs}
\author{M. Prunnila}
\thanks{Corresponding author \\
email: mika.prunnila@vtt.fi \\
Tel: +358 (0)40 537 8910 \\
Fax: +358 (0)20 722 7012}
\author{J. Ahopelto}
\affiliation{VTT Information Technology, P.O.Box 1208, FIN-02044 VTT, Finland }
\author{F. Gamiz}
\affiliation{Departamento de Eletr\'{o}nica y Tecnolog\'{\i}a de Computadores, Facultad
de Ciencias, Avenida Fuentenueva s/n 18071 Granada, Spain }

\begin{abstract}
We report on detailed room temperature and low temperature transport
properties of double-gate Si MOSFETs with the Si well thickness in the range 
$\sim 7-17$ nm. The devices were fabricated on silicon-on-insulator wafers
utilizing wafer bonding, which enabled us to use heavily doped metallic back
gate. We observe mobility enhancement effects at symmetric gate bias at room
temperature, which is the finger print of the volume inversion/accumulation
effect. An asymmetry in the mobility is detected at 300 K and at 1.6 K
between the top and back interfaces of the Si well, which is interpreted to
arise from different surface roughnesses of the interfaces. Low temperature
peak mobilities of the reported devices scale monotonically with Si well
thickness and the maximum low temperature mobility was 1.9 m$^{2}$/Vs, which
was measured from a 16.5 nm thick device. In the magneto transport data we
observe single and two sub-band Landau level filling factor behavior
depending on the well thickness and gate biasing. 

\end{abstract}

\maketitle

















\section{Introduction}

Advances in silicon-on-insulator (SOI) technology have made possible high
quality thin channel single-gate and double-gate (DG) Si
metal-oxide-field-effect-transistor (MOSFET) device structures \cite%
{celler:2003}, which are intensively explored at the moment\cite%
{shoji:1999,gamiz:2001b,esseni:2003,prunnila:2004b}. These both devices have
many advantages over the standard bulk Si MOSFETs. However, DG MOSFETs are
usually regarded as the most promising solution to the problems faced when
the device/gate length is down-scaled into sub-50 nm regime (short channel
effects) due to superior electrostatic gate control of the transistor
channel charge.\cite{celler:2003} In addition of boosting the gate control
the DG devices also provide other benefits in the form of enhanced electron
mobility.\cite{gamiz:2001b,esseni:2003,prunnila:2004b}

Apart from the relevance to the microelectronics industry the SOI material
has also enabled investigation of fundamental phenomena in SiO$_{2}$-Si-SiO$%
_{2}$ quantum well structures. Unfortunately, the electron mobility at the
Si-SiO$_{2}$ interface is intrinsically limited by the effects well known
from standard bulk Si-MOSFETs. Despite this fact, the strong electronic and
optical confinement provided by the SiO$_{2}$ barrier, many valley Si
conduction band and indirect gap makes this quantum well system a unique
tool to study several interesting effects in, for example, electron-hole
liquids \cite{pauc:2004} and bi-layer electron systems\cite%
{takashina:2004b,prunnila:2005}.

In this work, we report on the fabrication and detailed room temperature and
low temperature electronic properties of DG Si MOSFETs with Si well
thickness in the range $\sim 7-17$ nm. Mobility, electron density and high
magnetic field diagonal resistivity are mapped in large double-gate bias
windows, enabling detailed investigation of the transport properties at
different gate bias (electron distribution) symmetries.

\section{Experimental}

\begin{figure}[b]
\begin{center}
\includegraphics[width=55mm,height=!]{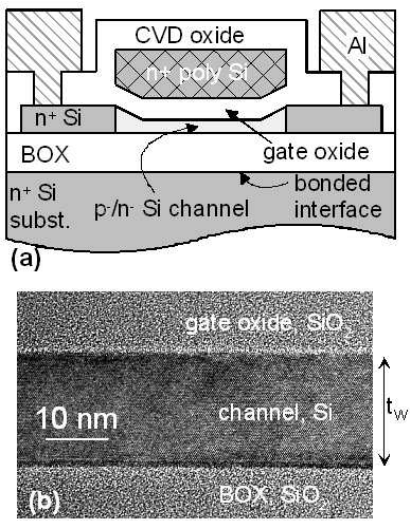}
\end{center}
\par
\caption{(a) Schematic illustration of the cross-sectional device structure
and (b) HRTEM image of Si channel ($t_{\text{W}}=18$ nm) of a DG MOSFET from
batch B.}
\label{TEM}
\end{figure}

The DG MOSFETs were fabricated on commercially available 100 mm unibond
(100) \text{SOI} wafers with n$^{-}$ (batch B) and p$^{-}$ (batch F) Si
layer . The nominal \text{Si} film thickness was 400 nm and the buried oxide
(BOX) was 400 nm thick. First, we exchanged the insulating handle wafer to a
n$^{+}$ wafer to enable efficient metallic back gating at all temperatures.
This procedure began by a growth of a 80 nm-thick dry oxide at 1000 $^{\circ
}$C. Then the SOI wafer was vacuum bonded to a n$^{+}$ (111) Si wafer with $%
\sim 2\times 10^{19}$ $\text{cm}^{-3}$ arsenic concentration. The bonded
interface was annealed at 1100 $^{\circ }$C and the original insulating $%
\sim $500 $\mu $m-thick handle wafer was removed by etching in 25\%
tetramethyl ammonium hydroxide solution at 80 $^{\circ }$C. Finally the
"old" BOX layer was stripped in a 10\% HF and as a result we had a SOI wafer
with heavily doped handle and 360 nm thick SOI film and 80 nm thick BOX
(back gate oxide).

The actual device fabrication for batch F begun by locally thinning the Si
layer in some parts of the wafer in order to fabricate devices with
different Si well thickness, $t_{\text{W}}$. This was done by utilizing
standard nitride masking and thermal oxidation, i.e., local oxidation of
silicon (LOCOS). Another two LOCOS\ steps were then used to define the thin
channels and active areas of the devices. In the channel thinning step the
Si thickness was reduced in the channel regions to create a recessed
source-drain MOSFET structures. To define the active areas the parts of the
Si layer that were not protected by the nitride mask were fully converted to
SiO$_{2}$. After oxide stripping and wafer cleaning a 40 nm-thick gate oxide
was grown at 1000 $^{\circ }$C in oxygen - DCE (dichloroethylene) ambient.
The applied DCE flow into the oxidation furnace corresponded to $\sim $2\%
of HCl. A 250 nm-thick polysilicon gate was deposited by CVD, implanted with
As and then patterned with UV-lithography and plasma etching. The contact
areas were implanted with As while the gate electrode protected the Si
channel. A $\sim $500 nm thick CVD oxide was deposited and the implanted
doses were activated at 950 $^{\circ }$C. Finally, after contact window
opening and Al metallization, the samples were annealed in H$_{2}$/N$_{2}$
ambient at 425 $^{\circ }$C for 30 min. Devices in batch B were fabricated
in a similar fashion. The major difference was that in this batch the active
areas of the devices were defined by etching through the Si layer instead of
LOCOS process. 
\begin{table}[t]
\caption{Device parameters: $t_{\text{W}}$ - Si well (channel) thickness,
type - type of the Si well, $\protect\mu _{\max }$ - maximum (or peak)
mobility, $T_{\protect\mu ,\max }$ - temperature where $\protect\mu _{\max }$
was measured.}
\label{dev_tab}
\begin{ruledtabular}
\begin{tabular}{ccccc}
Device & $t_{\text{W}}$(nm) & type & $\mu _{\max}$(m$^2$/Vs) & $T_{\mu,\max}$(K) \\ \hline
B-E721 & 16.5 & n$^{-}$ & 1.91 & 4.2 \\ 
B-E742 & 17.3 & n$^{-}$ & - & - \\ 
F-E741 & 6.8 & p$^{-}$ & - & - \\ 
F-E73 & 7.0 & p$^{-}$ & 0.76 & 1.6 \\ 
F-E42 & 14.2 & p$^{-}$ & 1.37 & 1.6 \\ 
\end{tabular}
\end{ruledtabular}
\end{table}

Figure \ref{TEM} shows a schematic cross-section of our DG MOSFET device
structure together with HRTEM image of a device from batch B. Properties of
the devices reported here are listed in Table \ref{dev_tab}. The gate oxide
thickness is $t_{\text{OX}}=$ 40 nm (43 nm) and the buried oxide thickness
is $t_{\text{BOX}}=$ 83 nm (80 nm) for batch F (B). The cited well type is
the Si layer type given by the SOI wafer manufacturer. No intentional doping
is introduced into the channel in order to maximize the low temperature
mobility. Further details about samples B-E721 and B-E742 can be found from 
\cite{prunnila:2005} and \cite{prunnila:2004b}, respectively.

All electrical characteristics reported here were obtained from Hall bar
structures \ with 100 $\times $ 1900 $\mu $m$^{2}$ channel dimensions and a
400 $\mu $m voltage probe distance. We used two different methods to
determine the electron density $N$:\ "split" capacitance-voltage (SCV)
method \cite{sodini:1982} at room temperature and Shubnikov-de Haas (SdH)
oscillations at low temperature. The mobility (or effective mobility) was
determined from $\mu =\sigma /eN$ where $\sigma $\ is the conductivity
measured by a four point method and $e$\ is the electronic charge. In the
room temperature SCV measurements we used Agilent 4294A precision impedance
analyzer at frequency of 691 Hz. This frequency was low enough to provide
results that were independent of the channel resistance when all the voltage
probes and the source and the drain were connected to the virtual ground of
the analyzer. The channel conductivity was determined by Agilent 4156 C
precision semiconductor parameter analyzer. DC offsets were systematically
removed from this data by applying at least four different source-drain bias
values at each gate voltage point. In the low temperature characterization
the samples were mounted to a sample holder of \ a He-3 cryostat and the
electrical measurements were performed utilizing standard low frequency
lock-in techniques. A combination of voltage and current excitation was
utilized in order to keep the source-drain bias from heating the electrons
above the lattice temperature at sub-1 K temperatures.

\section{Results and Discussion}

\subsection{Room Temperature Mobility}

\begin{figure*}[t]
\begin{center}
\includegraphics[width=164mm,height=!]{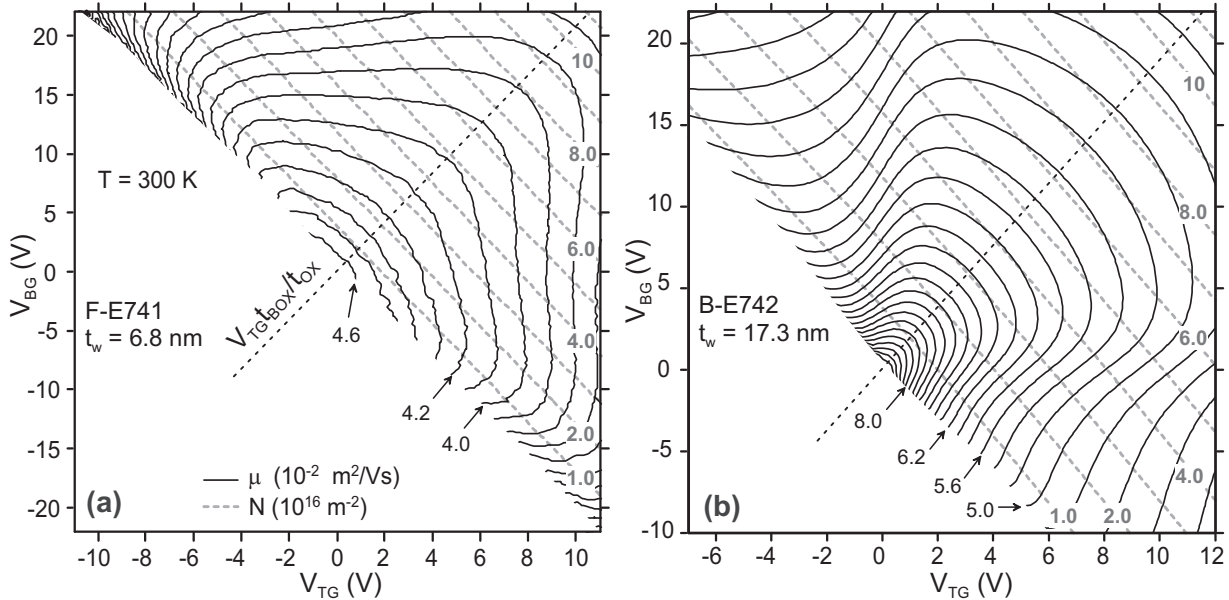}
\end{center}
\par
\caption{Effective electron mobility $\protect\mu $ (black thin curves) and
SCV electron density $N$ (gray thick dashed curves with gray bold labels) as
a function of top gate voltage $V_{\text{TG}}$ and back gate voltage $V_{%
\text{BG}}$ measured from (a) 6.8 nm thick and (b)\ 17.3 nm thick DG device
at 300 K. The contour spacing for $N$ is 1.0$\times $10$^{16}$m$^{-2}$. For $%
\protect\mu $ the spacing is 0.1 and 0.2$\times $10$^{-2}$ m$^{2}$/Vs in (a)
and (b), respectively. The dashed line is the symmetric gate bias line $V_{%
\text{BG}}=$ $V_{\text{TG}}t_{\text{BOX}}/$ $t_{\text{OX}}$. }
\label{muN_map}
\end{figure*}

\begin{figure}[b]
\begin{center}
\includegraphics[width=65mm,height=!]{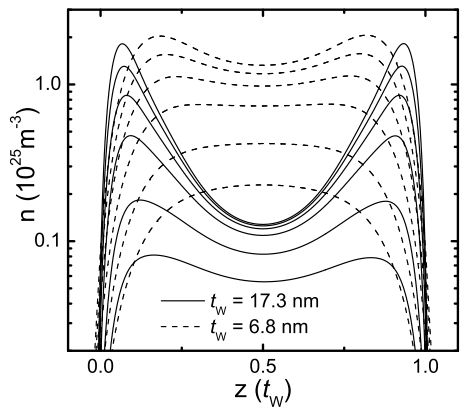}
\end{center}
\par
\caption{Self-consistently calculated quantum mechanical electron
distribution in the Si well at different electron density values (N =\
1,2,4,6,8,10$\times 10^{16}$ m$^{-2}$ from bottom to top) at symmetric gate
bias at 300 K. The calculations were performed for 12 sub-bands within the
Hartree approximation.}
\label{edist}
\end{figure}

Figure \ref{muN_map} shows experimental constant mobility and electron
density contours of two DG MOSFETs with $t_{\text{W}}=$ 6.8 nm (F-E741) and $%
t_{\text{W}}$ = 17.3 nm (B-E742) as a function of top gate voltage $V_{\text{
TG}}$ and back gate voltage $V_{\text{BG}}$ at 300 K. The family of curves
explicitly shows how the electron mobility behaves as a function of carrier
density and gate biases. We can see that for both devices on any of the
constant $N$ contours the mobility maximum occurs in the vicinity of the
symmetric gate bias line $V_{\text{TG}}/$ $t_{\text{OX}}=$ $V_{\text{BG}}/t_{%
\text{BOX}}$ [dashed black lines in Fig. \ref{muN_map}]. When gate bias
asymmetry is increased the mobility decreases monotonically. The mobility
enhancement towards the symmetric bias line can be related to the volume
inversion/accumulation effect, where the electron gas spreads through out
the whole Si well. This is illustrated by self-consistent Hartree electron
distributions in Fig. \ref{edist}.

The mobility modulation along the constant $N$ contours follows mainly from
the modulation of phonon scattering, surface roughness scattering, and
conduction effective mass.\cite{gamiz:2001b,gamiz:2003} The first two have
minimum and the latter maximum at symmetric bias. However, the phonon and
the surface roughness scattering have the strongest influence on the
mobility \cite{gamiz:2003} and this leads to the experimentally observed
behavior of Fig. \ref{muN_map}. The large difference between the magnitude
of the mobilities of \ the two devices follows from the Si well thickness
dependency of the effective mobility (not from sample quality) and it is
consistent with Monte Carlo simulations \cite{gamiz:2003} and experiments of
other groups \cite{esseni:2003}. 
\begin{figure}[t]
\includegraphics[width=90mm,height=!]{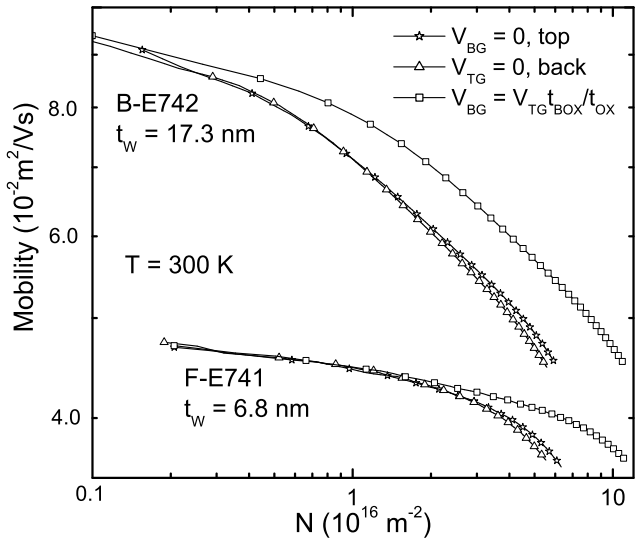} 
\caption{Top and back interface and symmetric gate bias effective electron
mobility as a function of SCV electron density at 300 K. }
\label{muvsN300K}
\end{figure}

It is evident from Fig. \ref{muN_map}(a) that the mobility maxima do not
occur precisely on the symmetric bias line for the thinner device F-E741 at
higher electron densities; the maxima are shifted towards larger (smaller) $%
V_{\text{TG}}$ ($V_{\text{BG}}$). More careful inspection reveals that the
behavior is actually similar for the thicker device B-E742. The observed
effect is brought out more clearly in Fig.\ref{muvsN300K}, which shows the
top interface ($V_{\text{BG}}=\ 0$), back interface ($V_{\text{TG}}=\ 0$),
and symmetric bias mobilities as a function of electron density . We can
observe that the back interface mobility falls below the top interface
mobility in both devices at high electron densities. As this top/back
interface mobility difference increases as a function of electron density we
interpret that the Si-BOX interface has larger surface roughness than the
Si-gate oxide interface.

\subsection{Magneto Transport}

\begin{figure*}[t]
\begin{center}
\includegraphics[width=152mm,height=!]{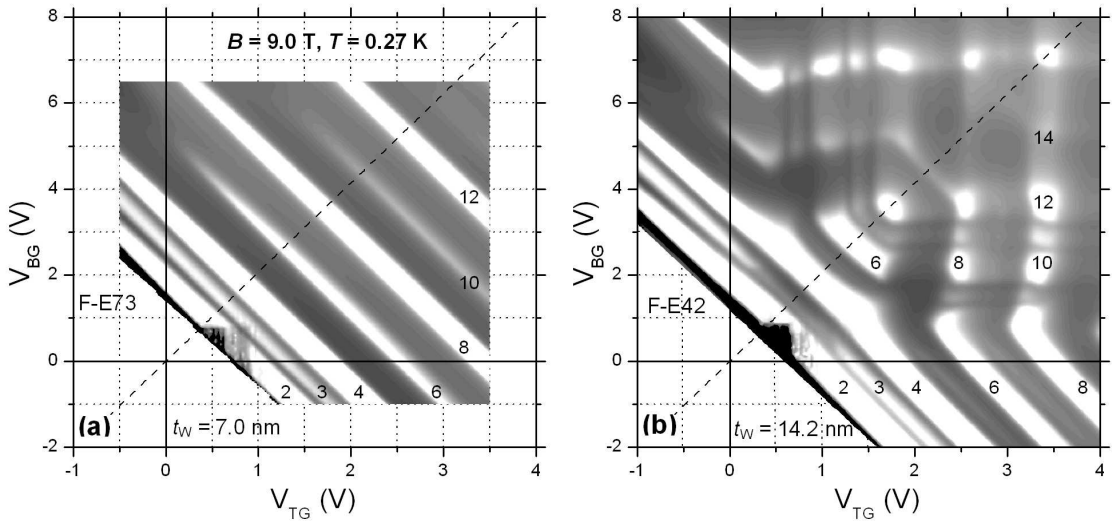}
\end{center}
\par
\caption{Gray scale plot of longitudinal resistivity $\protect\rho _{\text{%
xx }}$ of two DG MOSFETs at $B$ = 9.0 T at 0.27 K. The dashed line is the
symmetric gate bias line $V_{\text{BG}}=$ $V_{\text{TG}}t_{\text{BOX}}/$ $t_{%
\text{OX}}$. The gray scale is proportional to $\protect\rho _{\text{xx}}$:
dark and light color correspond to high and low resistance, respectively.
The regions showing the grid are excluded in the measurements. The numbers
inside the axis indicate the Landau level filling factors $\protect\nu $.
Only few $\protect\nu $ are indicated in (b) and missing values can be
simply obtained by counting the minima. }
\label{Rxx_maps}
\end{figure*}

Figure \ref{Rxx_maps} shows the diagonal resistivity $\rho _{xx}$ measured
as a function of top and back gate voltages from two DG MOSFETs F-E73 ($t_{%
\text{W}}=\ 7.0$ nm) and F-E42 ($t_{\text{W}}=\ 14.2$ nm) at $B$ =\ 9.0 T at
0.27 K. The numbers inside the axis indicate the Landau level (LL) filling
factors $\nu =Nh/eB$. The bright and dark color correspond to low and high $%
\rho _{xx}$, respectively. In the white regions $\rho _{xx}\sim 0$ and Hall
resistance $\rho _{xy}$ is equal to integer quantum Hall (QH) value $%
h/e^{2}\nu $. The obscured maximum slightly below the symmetric bias line
and close to the threshold in both $\rho _{xx}$ data is an experimental
artefact (present below $\sim 1$ K).\cite{note:arte} 
We first draw our attention to the LL filling factor behavior of the thin
sample F-E73 [Fig.\ref{Rxx_maps}(a)]. We can observe that the $\rho _{xx}$
minima are continuous trajectories, which suggest that only the ground 2D\
sub-band is populated. The constant LL trajectories $\nu =4(k+1)$, $\nu =2k+3
$ and $\nu =4k+2$ ($k=0,1,2...$) can be related to cyclotron, valley and
spin gaps, respectively. When the gate biases confine the electron gas
closer to the back interface ($V_{\text{BG}}>$ $V_{\text{TG}}t_{\text{BOX}}/$
$t_{\text{OX}}$) the $\rho _{xx}$ minima corresponding to\ $\nu =6$ and $\nu
=10$ become shallow. This weakening of the minima can be addressed to the
stronger elastic scattering in the vicinity of the Si-BOX interface, which
was already detected in the room temperature mobility (see also next
sub-section), leading to disorder broadening of the LLs. \ 

The thick sample F-E42 [Fig.\ref{Rxx_maps}(b)] behaves similarly in
comparison to F-E73 when electron density is low (small $\nu $) and also
when $V_{\text{BG}}\lesssim 0.5$ V or $V_{\text{TG}}\lesssim 0.25$ V. This
indicates single sub-band occupation and the $\rho _{xx}$ minimum
trajectories can be addressed to cyclotron, valley and spin gaps as was done
above. In the bias region where $V_{\text{BG}}\gtrsim 0.5$ V, $V_{\text{TG}
}\gtrsim 0.25$ V and $\nu >4$ the trajectories are broken into a 2D pattern,
which is a signature of two sub-band (bi-layer) transport \cite%
{muraki:1999,prunnila:2005,takashina:2004b}. By analyzing the SdH
oscillations of $\rho _{xx}$ as a function of inverse magnetic field (at 0.3
K) in the spirit of Ref. \cite{muraki:2000} along the symmetric bias line we
find that when top gate is adjusted between $V_{\text{TG}}=+0.9-+2.78$ V the
energy spacing of the bonding - antibonding sub-bands, $\Delta _{\text{BAB}
}, $ varies monotonically from $23$ to $2$ meV. \ 

The threshold for the second sub-band is already at $V_{\text{TG}}\approx
0.75$ V on the symmetric bias line, which is roughly on the $\nu =3$
trajectory in Fig. \ref{Rxx_maps}(b). Therefore, Zeeman and valley splitting
can push the lower valley spin down state of the second sub-band in to the
valley gap $\nu =3$ of the first sub-band around the symmetric bias, where $%
\Delta _{\text{BAB}}$ has a minimum value. This could explain the
destruction of the $\nu =3$ QH\ state (appearance of finite $\rho _{xx}$) at
symmetric gate bias. Note that the effect is complicated by the possible
symmetry dependency of valley splitting\cite{takashina:2004b} and also by
modulation of interface trap Coulomb scattering. These effects will be
further explored elsewhere.

\subsection{Low Temperature Mobility}

\begin{figure}[h]
\begin{center}
\includegraphics[width=80mm,height=!]{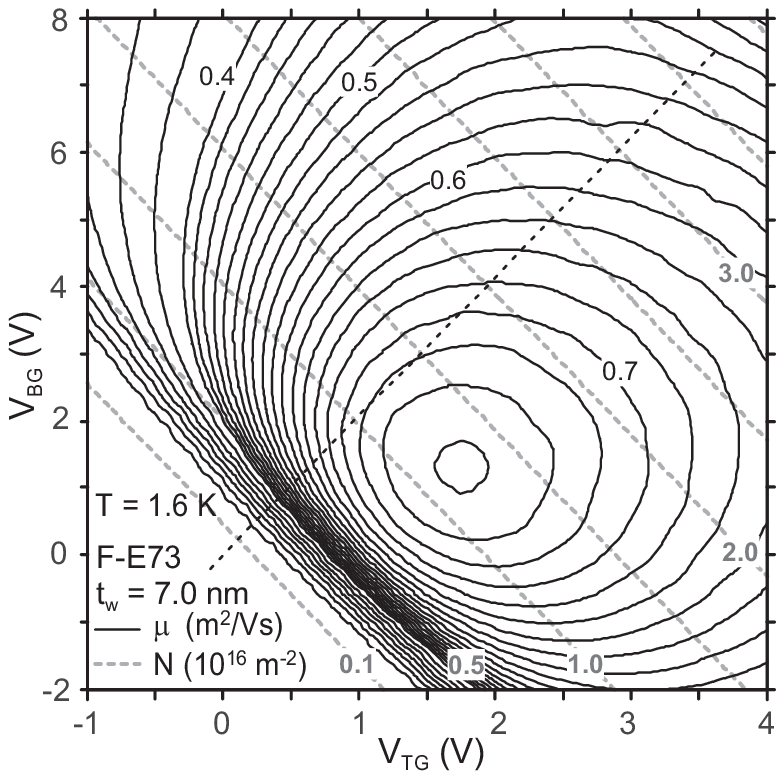}
\end{center}
\par
\caption{Contour plot of electron mobility $\protect\mu $ (black thin curves
with black labels) and SdH\ electron density $N$ (gray thick dashed curves
with gray bold labels) measured from the 7.0 nm thick DG device. Mobility is
determined from conductivity measured at 1.6 K and SdH\ electron density,
which is measured at $B$ = 2.5 T at 0.27 K. The contour spacing for $\protect%
\mu $ above (below) 0.4 m$^{2}$/Vs is 0.02 m$^{2}$/Vs (0.05 m$^{2}$/Vs). The
dashed diagonal line is the symmetric gate bias line. }
\label{muN_map1K}
\end{figure}
Figure \ref{muN_map1K} shows experimental constant mobility contours at 1.6
K and SdH electron density contours of \ device F-E73 as a function of $V_{%
\text{TG}}$ and $V_{\text{BG}}$. The electron density is determined from SdH
oscillations at 0.27 K: $N$ is obtained from similar constant $\nu $
trajectories in $\rho _{xx}$ at constant magnetic field that were discussed
above. Magnetic field value of $B$ =\ 2.5 T was chosen in order to keep all $%
\rho _{xx}$ minima clearly above zero, which enabled accurate determination
of constant LL trajectories. In the conductivity measurement at 1.6 K a
small magnetic field ($0.2$ T) was applied to suppress the quantum
corrections of conductivity.

The electron density and gate bias symmetry dependency of low temperature
mobility of F-E73 in Fig. \ref{muN_map1K} closely resembles that of room
temperature effective mobility of F-E741 [Fig. \ref{muN_map}(a)]. Due to
absence of phonon scattering the asymmetry in the elastic scattering
properties of the two interfaces is now brought out more clearly. It is
evident that the maximum mobility is shifted from the symmetric gate bias
position towards larger $V_{\text{TG}}$ and smaller $V_{\text{BG}}$ on any
of the constant $N$ contours.

If we adjust the gate voltages along an arbitrarily chosen straight line
(direction) in Fig. \ref{muN_map1K} in such a fashion that $N$ increases the
mobility shows the typical MOSFET behavior, where the mobility first
increases at low carrier density, reaches a maximum value and then decreases
at high carrier density. This behavior can be addressed to well known
elastic scattering mechanisms: Coulomb scattering that is most effective at
low $N$ and interface roughness scattering that is most effective at high $N$
.\cite{ando:1982}

When $t_{\text{W}}\gtrsim 10$ nm the second sub-band becomes populated even
when both gates have a modest positive bias, as was demonstrated in the
previous sub-section. \ This alters the mobility behavior and in general the
simple Coulomb/surface roughness-scattering picture is lost. \ Further
effects arises from the fact that in some bias ranges the electrons in the
second sub-band can be localized \cite{prunnila:2005}. The co-existence of
localized and non-localized electrons complicates the scattering mechanisms 
\cite{popovic:1997,feng:1999} and discussion of such effects is beyond the
scope of this paper. Therefore, full $V_{\text{TG}}-V_{\text{BG}}$ mobility
dependency of the devices with $t_{\text{W}}\gtrsim 10$ nm will be reported
elsewhere. Here we only cite the maximum mobilities, which can be found from
Table \ref{dev_tab}. The maximum mobility $\mu _{\max }$ decreases with
decreasing $t_{\text{W}}$ as expected. $\mu _{\max }$ scaling with $t_{\text{
W}}$ is consistent with recent experimental observations on single gate SOI
devices\cite{prunnila:2004}.

Finally, we note that as the maximum mobilities for all devices are
relatively high it is unlikely that the mobility degradation at the back
interface could originate from increased Coulomb scattering; the degradation
is mainly due to interface roughness scattering. High mobility also
indicates extremely low interface trap density, which justifies our
assumption that in the room temperature measurements the SCV electron
density corresponds to that of mobile electrons. \ 

\section{Summary}

In summary, we have reported on detailed room temperature and low
temperature transport properties of ultrathin channel double-gate Si
MOSFETs. The devices were fabricated on SOI wafers utilizing wafer bonding,
which enabled us to use heavily doped metallic back gate. The devices showed
mobility enhancement effects at symmetric gate bias at room temperature,
which is the finger print of the volume inversion/accumulation effect. Small
asymmetry in the mobility could be detected at 300 K between the top and
back interfaces of the Si well. The effect could be enhanced at low
temperatures and the mobility asymmetry was interpreted to arise from
different interface roughness of the top and back interface. Low temperature
peak mobilities of the reported devices scale monotonically with Si well
thickness and the maximum low temperature mobility was 1.9 m$^{2}$/Vs, which
was measured from a 16.5 nm thick device. From the magneto transport data we
observed single and two sub-band transport effects depending on the well
thickness and gate biasing.

\section{Acknowledgements}

Technical assistance by M. Markkanen in the sample fabrication is gratefully
acknowledged. K. Henttinen is thanked for the wafer bonding and S. Newcomb
is thanked for the TEM analysis. This work has been partially funded by EU
(IST-2001-38937 EXTRA), by the Academy of Finland (\# 205467 CODE) and GETA
graduate school.



\bibliographystyle{ieeetr}
\bibliography{SSE05_refs}

\begin{thebibliography}{10}
\bibitem{note:arte}
At low electron densities close to the symmetric bias line $\rho_{xx}$ shows
  gate voltage dependent reproducible fluctuations, which are present roughly
  below 1 K. These probably arise from the tapered Si well regions between the
  Hall bar voltage probes and n$^+$ contacts. The mobility in these regions is
  expected to be extremely low due to undefined crystal orientation of the Si -
  gate oxide interface.
\end{thebibliography}




\end{document}